\documentclass{emulateapj}

\usepackage{float}
\usepackage{epsfig,floatflt}
\usepackage{subfigure}

\begin{document}

\title{Evidence of vorticity and shear at large angular scales 
in the \emph{WMAP} data: a violation of cosmological isotropy?}

\author{T.\ R. Jaffe\altaffilmark{1}, 
A.\ J.\ Banday\altaffilmark{1}, H.\ K.\ Eriksen\altaffilmark{2}, K.\ M.\ G\'orski\altaffilmark{3,4},  F. K. Hansen\altaffilmark{2}
}

\altaffiltext{1}{Max-Planck-Institut f\"ur Astrophysik,
Karl-Schwarzschild-Str.\ 1, Postfach 1317, D-85741 Garching bei
M\"unchen, Germany; tjaffe@MPA-Garching.MPG.DE, 
banday@MPA-Garching.MPG.DE.}


\altaffiltext{2}{Institute of Theoretical Astrophysics, University of
Oslo, P.O.\ Box 1029 Blindern, N-0315 Oslo, Norway; 
h.k.k.eriksen@astro.uio.no, f.k.hansen@astro.uio.no.} 

\altaffiltext{3}{JPL, M/S 169/327, 4800 Oak Grove Drive, Pasadena CA
91109; Krzysztof.M.Gorski@jpl.nasa.gov}

\altaffiltext{4}{Warsaw University Observatory, Aleje Ujazdowskie 4, 00-478
Warszawa, Poland}


\begin{abstract}
Motivated by the large-scale asymmetry observed in the cosmic
microwave background sky, we consider a specific class of anisotropic
cosmological models -- Bianchi type VII$_h$ -- and compare them to the
\emph{WMAP} first-year data on large angular scales.
Remarkably, we find 
evidence of a correlation which is ruled out
as a chance alignment at the $3\sigma$ level.  The best fit
Bianchi model 
corresponds to $x=0.55$, $\Omega_0=0.5$, a rotation axis in the   
direction $(l,b)=(222\degr,-62\degr)$, 
shear $\left ( \frac{\sigma}{H} \right )_0=2.4\times 10^{-10}$ and a 
right--handed vorticity $\left ( \frac{\omega}{H}\right )_0=6.1\times10^{-10}$.
Correcting for this component 
greatly reduces the significance of the large-scale power asymmetry,
resolves several anomalies detected on large angular scales
(ie. the low quadrupole amplitude and quadrupole/octopole planarity and alignment),
and can account for a non--Gaussian ``cold spot''
on the sky.
Despite the apparent inconsistency with the best-fit parameters 
required in inflationary models to account for the acoustic peaks,
we consider the results sufficiently provocative to merit
further consideration.
\end{abstract}


\maketitle

\section{Introduction}
\label{sec:introduction}

Despite initial reactions to the first year 
\emph{WMAP}\footnote{Wilkinson Microwave Anisotropy Probe}
data -- ``the most revolutionary result is that there are no revolutionary results''
\citep{bahcall:2003} -- continued assessment has revealed some
interesting discrepancies with the inferred best-fit, 
cosmological constant dominated, inflationary model of the
primordial fluctuation spectrum.

A debate about the apparently low quadrupole amplitude
(originally observed by \emph{COBE}-DMR) 
has arisen 
\citep{efstathiou:2004} along with claims about other anomalies on
large angular scales. 
\citet{de Oliveira-Costa:2004} demonstrated a curious planarity and alignment
of the quadrupole and octopole. 
In addition,
\citet{vielva:2004} and \citet{cruz:2005} have detected a localised
source of non-Gaussianity, in the form of a very cold spot on the sky
of angular scale $\sim 10^{\circ}$.  
Of particular interest to us, however, 
was the discovery of a remarkable asymmetry in large-scale
power as measured in the two hemispheres of a particular
reference frame.    
\citep{eriksen:2004a,hansen:2004a,hansen:2004b}.  
It remains unclear why this was not detected by
the bipolar power spectrum of \cite{hajian:2005}.
Nevertheless, this has provided the motivation to investigate
anisotropic cosmological models.  

Following \cite{bunn:1996} and \cite{kogut:1997}, 
we focus on a specific class of models -- Bianchi type
VII$_h$ \citep{barrow:1985}.
Such models were previously compared to the \emph{COBE}-DMR data in
\cite{kogut:1997}, where limits 
on the shear $\left (\frac{\sigma}{H} \right )_0 < 10^{-9}$ and 
vorticity $\left (\frac{\omega}{H}\right )_0 < 6\times10^{-8}$ 
were established.
We consider that the observed anisotropy is the sum of two contributions
-- an \lq isotropic' term which is connected to variations in the density
and gravitational potential, and a term from the anisotropic metric.
A mechanism to generate the isotropic fluctuations is required,
particularly on small angular scales where the Bianchi contribution 
is negligible, but this need not be inflationary. In our analysis 
we smooth the data to probe only the low-$\ell$ regime, 
and find that results are practically insensitive 
to the choice of spectrum for the isotropic component 
on large-angular scales.

Remarkably, we find a statistically significant
correlation between one of these models and the \emph{WMAP} data.
Such a result may help to resolve some of the more unusual observed features
of the microwave sky, 
albeit at the introduction of a new conundrum --
the large-scale anisotropy is described at least in part by a
low density Bianchi model, 
whereas the smaller scale fluctuations 
are consistent with a cosmological constant dominated, critical
density model in an inflationary scenario.



\section{Data, simulations and templates}
\label{sec:data}

In this analysis, we utilize the first-year \emph{WMAP}
data\footnote{available at http://lambda.gsfc.nasa.gov}
in both raw and foreground template corrected forms
\citep{bennett:2003a,bennett:2003b}
and three heavily processed maps derived to minimize the
foreground contribution in a template independent way
(WILC -- \citealt{bennett:2003b}; LILC -- \citealt{eriksen:2004b}; 
TOH -- \citealt{tegmark:2003}).
For the three ILC-like maps we consider the full sky in the analysis, 
whereas the Kp2 or Kp0 masks \citep{bennett:2003b} are
imposed otherwise.
All are smoothed to a common resolution of $5.5\degr$ FWHM and
downgraded to a HEALPix\footnote{http://www.eso.org/science/healpix/} 
resolution of $N_{\textrm{side}}=32$ corresponding to harmonics up to
$\ell=64$. Such processing does not compromise the analysis since the
Bianchi models exhibit little power outside of this $\ell$-range.

To assess the statistical significance of the fits we compare our
results to an ensemble of 10\,000 LILC simulations, produced using the
pipeline described by \citet{eriksen:2004b}.

The Bianchi model templates are constructed using the formalism of
\citet{barrow:1985}.
These models are parameterized by $\Omega_0$, $ x =
\sqrt{h/(1-\Omega_0)} $ (where $h$ is the scale on which basis vectors
change orientation), and a handedness.  These models have a
prefered axis along which the expansion rate is different, and about
which the basis vectors themselves rotate.  The smaller the $x$, the
more rotation in a given distance traversed along a geodesic, and
therefore the tighter the observed spiral pattern of induced
temperature anisotropies.  The smaller the $\Omega_0$, the larger is
the asymmetry along the axis and the more focused is the structure in
only one direction.  In this
analysis, we consider the range $0.1\le x \le 10$ and
$0.1\le\Omega_0\le 1$.  The resulting temperature anisotropy pattern
is described by Equation 4.11 of
\citet{barrow:1985}.  The fit amplitude determines the shear
$\left(\frac{\sigma}{H}\right)_0$ and the vorticity
$\left(\frac{\omega}{H}\right)_0$ following Equation 4.8.

\section{Method}
\label{sec:computations}

In order to compute the correlation between the data $\mathbf{d}$
and a template $\mathbf{t}$, we adopt the formalism of
\citet{gorski:1996} and \citet{kogut:1997} . Specifically, at a given
frequency we estimate
a template coupling constant $\alpha$ by minimizing
$\chi^{2} = (\mathbf{d}-\alpha\mathbf{t})^\textrm{T} \,\mathbf{M}^{-1}
\, (\mathbf{d}-\alpha\mathbf{t})$,
where $\mathbf{M}$ is the CMB signal plus noise covariance matrix.
The solution to this problem (and corresponding uncertainty) is
$\alpha = ( {\bf t}^T\,{\bf M}^{-1}\, {\bf d} )/( {\bf t}^T\,{\bf
M}^{-1} \, {\bf t} )$ and  $\delta\alpha=({\bf t}^T\,{\bf
M}^{-1}\, {\bf t} )^{-1/2}$.
The generalization to the multi-frequency case can be found in
\citet{gorski:1996}.
In what follows, the CMB signal covariance is specified by the \emph{WMAP}
best-fit theoretical power spectrum.

In our analysis, we must determine $\alpha$ over all possible models
and, for each model, all relative rotations between the corresponding
template and the data.  Such an analysis is generally computationally
demanding, but can be accelerated by a full-sky harmonic-space approach.
In this case, if 
the noise covariance
is a diagonal approximation based on the mean noise level
for the map under 
consideration\footnote{This approximation has a negligible 
effect on results at the \emph{WMAP} signal-to-noise ratio.}
then the denominator of $\alpha$ 
is invariant under template rotation
and simply computed for any given model. The numerator can then
be evaluated from the convolution of the template and the ``whitened''
data map ${\bf M}^{-1} {\bf d}$.  This operation is performed
using the total convolution method of \citet{wandelt:2001} to
efficiently evaluate all such convolutions over all possible
rotations.  

The technique is only straightforward when considering
full sky coverage. Thus, despite concerns about the effectiveness 
of foreground removal in the Galactic plane, we analyse 
the foreground-cleaned LILC map.
Since chance alignments between the Bianchi template and the CMB anisotropy
arising even in the absence of rotation and shear can generate 
non-zero correlations, we follow \citet{kogut:1997} and 
define the parameter $\Gamma=\alpha/\delta\alpha$ to help
reject this possibility at a given confidence and to select the best model.
A reference ensemble of 10\,000 LILC simulations were analysed using
the above formalism to establish the distribution of $\alpha$
values and hence to estimate the significance of our best fit model.

Once the model and orientation axes are specified, we apply a pixel-based
analysis to various combinations of the \emph{WMAP} sky maps
and impose the conservative Kp0 Galactic cut to reduce foreground contamination,
in order to confirm these results.
In the case of incomplete sky analysis, the amount of template
structure which is masked can affect the uncertainty of the fit. 
Nevertheless, for the best fit orientation where the structure is
largely outside the excluded Galactic Plane region, we find that the
fit and its uncertainty are dominated by the CMB covariance rather  
than the particulars of the noise properties or sky cut.
Therefore, although we calibrate
using simulations strictly only applicable to the LILC result, we
expect that the predicted levels of chance alignment detections should
be quite robust.

\section{Results}
\label{sec:results}

\begin{figure}
\begin{center}
\mbox{\epsfig{figure=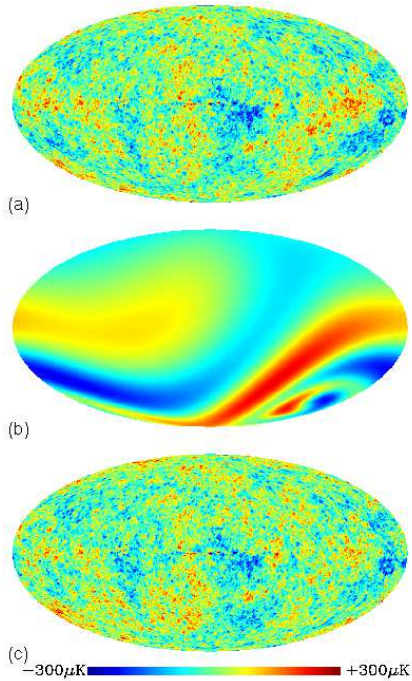,width=0.65\linewidth,clip=}}
\end{center}
\caption{Top panel: \emph{WMAP} Internal Linear Combination map. Middle
  panel: Best-fit Bianchi VII$_h$ template (enhanced by a factor of
  four to show the structure). Bottom panel: Difference, {\it i.e.} the ``Bianchi-corrected'' ILC map}
\label{fig:maps}
\end{figure}

The results from the analysis are summarized in Table
\ref{tab:amplitudes} for our best fit model at $x=0.55$ and
$\Omega_0=0.5$, right--handed.  The excess power in the southern
hemisphere requires a model with low $\Omega_0$ to focus the spiral.
The cold spot and surrounding structure effectively select the
particular $x$.  See Figure \ref{fig:maps}.
  The best-fit Euler angles\footnote{We adopt
the zyz-convention of
\citet{wandelt:2001} for the three Euler angles.}  for the WILC map
were found to be $(\Phi_2,\Theta,\Phi_1) = (42\degr, 28\degr,
-50\degr)$, and in what follows this is adopted as the best-fit
axis. However, both the LILC and TOH sky maps as analysed on the
full-sky yielded statistically consistent best-fit axes.  This
rotation places the center of the spiral structure, originally at the
$-z$ axis, at $(l,b)=(222\degr,-62)$.  For each map, we tabulate the
sheer and vorticity, and corresponding estimate of significance as
computed from the fraction of LILC simulations with a smaller $\alpha$
value than the observed map.  The estimated shear amplitudes are
consistent with a value $\sim 2.4\times10^{-10}$, significant at the
99.8\% confidence level relative to the Monte Carlo ensemble.

\begin{deluxetable}{lcccc}
\tablecaption{Fitted template amplitudes\label{tab:amplitudes}} 
\tablecomments{Amplitudes of the best fit Bianchi model.  
See text.  $^a$ Fraction of LILC simulations with lower amplitudes.
$^b$ Including simultaneously derived foreground corrections.  $^c$
Using \emph{WMAP} frequency maps pre-corrected for Galactic
foregrounds.  }
\tablewidth{0pt}
\tablecolumns{6}
\tablehead{  & $\left(\frac{\sigma}{H}\right)_0$ & $\left(\frac{\omega}{H}\right)_0$ &
   &  \\
Map  & ${\scriptstyle (\times10^{-10})}$ & ${\scriptstyle (\times10^{-10})}$ &
 $P(\alpha_{\textrm{sim}} < \alpha_{\textrm{obs}})\tablenotemark{a}$}
\startdata
WILC   & $2.39 {\pm 0.45}$ & $6.14$ &  $99.7\%$ \\ 
LILC  & $2.37 {\pm 0.45}$ & $6.07$ &  $99.6\%$ \\
TOH   & $2.25 {\pm 0.45}$ & $5.76$ &  $98.6\%$ \\

K\tablenotemark{b}   & $2.39 {\pm 0.46}$ & $6.14$   & $99.7\%$\\
Ka\tablenotemark{b}  & $2.35 {\pm 0.46}$ & $6.04$   & $99.5\%$\\
Q\tablenotemark{b}   & $2.35 {\pm 0.46}$ & $6.04$   & $99.5\%$\\
V\tablenotemark{b}   & $2.42 {\pm 0.46}$ & $6.22$   & $99.8\%$\\
W\tablenotemark{b}   & $2.48 {\pm 0.46}$ & $6.38$   & $99.9\%$\\

Q+V+W\tablenotemark{bc} & $2.42 {\pm 0.46}$           & $6.22$    & $99.8\%$\\
V+W\tablenotemark{bc}   & $2.43 {\pm 0.46}$           & $6.25$    & $99.8\%$\\
Q-V\tablenotemark{bc} &  $-0.07\pm 0.01$ &  0.17 &   - & \\
V-W\tablenotemark{bc} &  $-0.06\pm 0.01$ &  0.14 &   - & \\
Q-W\tablenotemark{bc} &  $-0.12\pm 0.01$ &  0.31 &   - & \\

\enddata

\end{deluxetable}

In order to assess the effect of residual foreground
contamination (ie. due to signals not well-traced by the foreground
templates) several difference maps were constructed from the
\emph{WMAP} foreground corrected maps. The results indicate that
foreground residuals are unlikely to result in spurious detections.

In Figure \ref{fig:maps} we show (from top to bottom) the WILC map,
the best-fit Bianchi model, and the difference between the two. 
It should be apparent that the ``Bianchi corrected'' map exhibits
greater isotropy than the WILC data.

In Figure \ref{fig:spectrum} we compare the power spectra of the
original and the Bianchi-corrected V+W linear combination 
map\footnote{The \emph{WMAP} team \citep{hinshaw:2003} use data solely 
from the V- and W-bands 
over the spectral range $\ell < 100$.}, 
as computed by the MASTER algorithm \citep{hivon:2002} for the Kp2
sky coverage. 
It should be noted that, whilst the 
quadrupole amplitude is significantly increased,
on average the Bianchi-corrected map has slightly less power on
large scales, although both corrected and uncorrected data are 
in good agreement for $\ell > 15$.
More importantly, the general shape of the Bianchi-corrected spectrum
is flatter than for the \emph{WMAP} best-fit power spectrum,
and in remarkable agreement with the theoretical fit made
by \citet{hansen:2004a} to the northern hemisphere data
(defined in the reference frame which maximises the power asymmetry)
which favours a lower value for $\tau$.

\section{Implications}
\label{sec:implications}



\begin{figure}
\mbox{\epsfig{figure=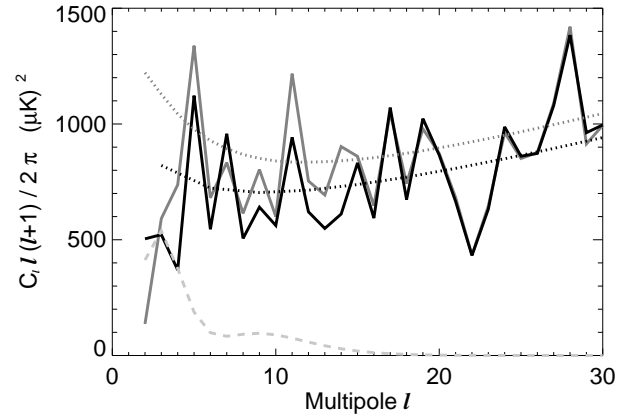,width=\linewidth,clip=}}
\caption{Comparison of power spectra. The gray and black solid lines show
the power spectrum estimated from the co-added V+W map before and
after correcting for the Bianchi template, respectively.  The
dotted gray and black lines shows the theoretical best-fit
power-spectra 
from the \emph{WMAP}-team analysis and \citet{hansen:2004a}
respectively. The latter is a fit to northern hemisphere
data alone. 
The light gray dashed line corresponds to the full
sky power spectrum of the best-fit Bianchi template.
}
\label{fig:spectrum}
\end{figure}

\noindent {\bf Quadrupole amplitude}: 
The \emph{WMAP} team suggest that the quadrupole
amplitude is significantly low, although 
other analyses have found it to be quite acceptable 
(Slosar \& Seljak 2004; O'Dwyer et al. 2004).
As seen in Figure \ref{fig:spectrum},
the Bianchi-corrected V+W map has a quadrupole amplitude of
$504\,\mu\textrm{K}^2$, compared to the uncorrected
amplitude of $137\,\mu\textrm{K}^2$ and the 
\emph{WMAP} theoretical best-fit spectrum 
value of $869\,\mu\textrm{K}^2$.   In this context, the
quadrupole amplitude should no longer be considered anomalous.
Whether the amplitude enhancement itself 
requires an unusual cancelation between the
intrinsic and Bianchi--induced quadrupoles, 
which could also be considered a \lq fine-tuning'of the model,
is deferred to a later analysis.

\noindent {\bf Low-$\ell$ anomalies}: 
An alignment between the quadrupole and the octopole has been claimed
by  \citet{de Oliveira-Costa:2004} and \citet{copi:2004}.
Quantitative calculations similar to those of \citet{de
Oliveira-Costa:2004} and \citet{eriksen:2004b} show that a stronger
planarity is expected by chance with a probability of 52\% for both
the $\ell=2$ and 3 modes after subtracting the Bianchi
template. The angle
between the preferred directions of the $\ell=2$ and 3 modes is
$70\degr$ after subtracting the Bianchi template, compared to $12\degr$
before.
Additionally, the $\ell=5$ and 6 modes (Eriksen et
al. 2004b) become less anomalous, with significances dropping 
to $1.5\sigma$ in both cases.



\noindent {\bf Large-scale power asymmetry}: \citet{eriksen:2004a} reported that the large-scale
power ($\ell \lesssim 40$) in the \emph{WMAP} data is anisotropically
distributed over two opposing hemispheres, with a significance of
$3\sigma$ compared with simulations. 
Repeating the analysis and adopting the Kp2 sky coverage, 
we compare the corrected V+W \emph{WMAP} map 
with 2048 simulations. We find that 13.6\% of the simulations
have a larger maximum power asymmetry ratio than the Bianchi-corrected
map, whereas only 0.7\% have a larger ratio than the uncorrected
data. 
It is apparent that the maximum
power ratio between any two hemispheres is significantly suppressed after
subtracting the Bianchi template -- no asymmetry axis is found
at any statistically significant level.



\begin{figure}
\mbox{\epsfig{figure=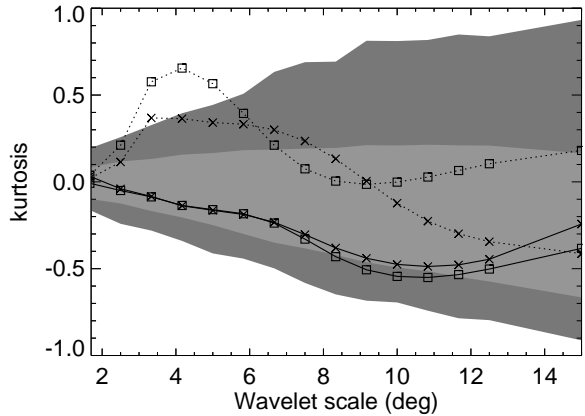,width=\linewidth,clip=}}
\caption{The kurtosis of the wavelet coefficients as a function of
wavelet scale.  The solid (dotted) lines show the results for the
northern (southern) Galactic hemisphere (excluding the region where
$|b|<20\degr$) computed from the \emph{WMAP} ILC map. The squares
(crosses) show the kurtosis before (after) correction for the Bianchi
template. Grey bands indicate the two-sided $68\%$ and $95\%$
confidence intervals obtained from 1000 Monte-Carlo simulations.  }
\label{fig:kurtosis}
\end{figure}

\noindent {\bf Wavelet kurtosis}: \citet{vielva:2004} used a
wavelet technique to detect an unusually cold spot 
($\sim 3\sigma$ significance relative to Gaussian simulations)
at Galactic coordinates $(l,b) = (207^{\circ},-59^{\circ})$. 
We repeat the analysis of \citet{vielva:2004}, and compute the
kurtosis of the wavelet coefficents as a function of scale from both
the WILC and the corresponding Bianchi-corrected map. For
computational convenience, a Galactic cut is imposed to exclude the
region where $|b|<20\degr$.  The results from this exercise are
reported in Figure
\ref{fig:kurtosis}.
After subtracting the Bianchi template, 
the southern hemisphere results are consistent with
the 95\% confidence intervals obtained from simulations --
no non-Gaussian features are apparent.

\section{Conclusions}
\label{sec:conclusions}

We have considered the Bianchi type VII$_h$ class of anisotropic
cosmological models and fitted the predicted CMB temperature
anisotropy patterns to the first year \emph{WMAP} data.  
The results are essentially independent of frequency or the
method of foreground subtraction.
A particular model with a rotation axis in the direction
$(l,b)=(222\degr,-62\degr)$, shear $\left ( \frac{\sigma}{H} \right
)_0=2.4\times 10^{-10}$ and a right--handed vorticity $\left (
\frac{\omega}{H}\right )_0=6.1\times10^{-10}$ yields the  best fit
to the data.   
An ongoing analysis will
improve the accuracy and completeness of the search of the model
space and examine other possibly significant models.

Should this result be considered more than a curiosity?  A skeptic
would no doubt stress the inconsistency between those cosmological
parameters describing the Bianchi anisotropy pattern (an open model
with $\Omega_0=0.5$) and the \emph{WMAP} best-fit cosmological
power-spectrum which must account for the acoustic peaks.  Yet
paradoxically, it is precisely the low value for $\Omega_0$ which is
required to allow the focussing of the Bianchi anisotropy pattern into
one hemisphere.  
The preferred Bianchi model reconciles the observed
asymmetric distribution of power on large angular scales, disrupts the
observed planarity and alignment of the quadrupole and octopole
moments, and provides an explanation for a highly non-Gaussian
signature on the sky -- unexpected results given that the model was
not selected based on these criteria, but solely on a statistical fit
of the predicted anisotropy pattern to the
\emph{WMAP} data.  Furthermore, the model is also consistent with the
tentative result from \citet{hansen:2004a} that the estimated optical
depth of $\tau=0.17$ on the (nearly) full sky \citep{kogut:2003} could
in large part originate in structure associated with the southern
hemisphere (in the reference frame which maximizes the power
asymmetry). 


While the consistency of our result on large angular scales with
models for the small-scale anisotropy remains ambiguous, 
we note that the models considered
include no dark energy contribution.  
A more self-consistent approach is clearly warranted, but we propose
that the result is sufficiently provocative to encourage a renewed exploration
of anisotropic models in a broader cosmological parameter
space. 

\begin{acknowledgements}
We are grateful to M.~Demianski, S.~Hervik, P. Mazur, and S. D. M.~White for
useful discussions.  H. K. E. acknowledges financial support from the
Research Council of Norway, including a Ph.\ D. scholarship. We
acknowledge use of the HEALPix software 
\citep{gorski:2005}
and of the Legacy Archive for Microwave Background
Data Analysis (LAMBDA). 
\end{acknowledgements}


\end{document}